\documentclass[a4paper,11pt]{article}
\pdfoutput=1 

\usepackage{jcappub} 

\usepackage[T1]{fontenc} 

\title{\boldmath The cosmology in f(R,T) gravity without dark energy}

\author{Gang Sun,}
\author{Yong-Chang Huang}

\affiliation{Institute of Theoretical Physics, Beijing University of Technology, Beijing 100136, China}

\emailAdd{sungang@bjut.edu.cn}
\emailAdd{ychuang@bjut.edu.cn}

\abstract{We examine the evolution of universe in $f(R,T)$ gravity where $R$ is the Ricci curvature and $T$ is the trace of energy momentum tensor.  We focus on the specific form $f(R,T)=R+f(T)$ and explicitly find out the relation between $f(T)$ and state of equation $w$ by adiabatic condition.  All possible modes of evolution of the scale factor for the five kinds of state of equation, $w=\pm1, \pm\frac{1}{3}, 0$ are exhibited. We simulate numerically the plots of redshift versus distance modulus for three interesting modes and then obtain the good fitting curve compared with the astronomical observation data.}

\begin{document}
\maketitle
\flushbottom

\section{Introduction}

The data of observation of Type Ia supernova suggested that our universe is expanding at an acceleration rate\cite{Perl,Riess98,Riess99}, which triggered the challenge for theoretical physics. People commonly call the unknown mechanism of causing the accelerated expansion dark energy. In GR, the simplest idea to cause the accelerated expansion is to introduce the cosmological constant. However, due to the fine-tuning problem, people turn to look for other programs of dynamics. More generally, a quintessence field\cite{Pee, Zla}, a scalar field carrying the kinetic with positive pressure and the potential with negative pressure, could be a candidate of dark energy. Unlike a cosmological constant, the quintessence is dynamic so that energy density and pressure are varying with time. It satisfies the equation of state (EOS) with $-1< w<0$. The more extreme scalar field carrying the EOS with $w<-1$ called phantom can also support accelerated expansion but it could accelerate too fast to result in the Big Rip\cite{Cald}.


On the other hand, the modified gravity could server as the candidate of dark energy to account for the accelerated expansion of the universe. The well-studied $f(R)$ gravity could generate late time acceleration by terms containing $1/R$ or a more term of positive curvature like $R^2$ added to Einstein Hilbert Lagrangian. \cite{Carr, Noj03,Anto, Sot} However, they could be incompatible with astronomical observation\cite{LiMiao}. Such as $f(\tau)$ gravity, to replace curvature $R$ with torsion $\tau$, it could also provide a mechanism of late time acceleration\cite{Bao, Hehl, Ho}. The cosmology based on the generalized scalar-tensor theory\cite{Ame}, the Gauss-Bonnet gravity\cite{Noj05} and quantum cosmology\cite{Hartle, Has, Ali} have been considered.

These different programs could describe the accelerated expansion of the universe and other issues to some extent, but none of them could solve all problems. Hence to investigate a new kind of gravity and its application is worth doing.
We will explore the cosmology in the $f(R,T)$ gravity\cite{Harko}, whose lagrangian is constructed by the Ricci curvature $R$ and the trace of energy-momentum tensor $T$. It can be regarded as the generalization of Brans-Dicke gravity. Some works on cosmology have been developed in the framework of $f(R,T)$ gravity\cite{Hou1, Hou2, Mub, Sub, Ham}.

The paper is organized as follow. In the section 2, introduce the basic formulae of the $f(R,T)$ model and the related equations for the specific form of $f(R,t)$. In the section 3, the FLRW field equations will be obtained. Furthermore, we will utilize the assumption of adiabatic process to produce the enough conditions to close the equations. In the section 4, we will examine some interesting cases by numerical simulation. In the final section, we will give a short comment on this paper.

\section{$f(R,T)$ gravity}
\subsection{The variation and field equations}

Let's start from the the general setting of $f(R,T)$ theory. In this framework, the Lagrangian reads\cite{Harko}
\begin{eqnarray}\label{1.1}
    I=\frac{1}{2\kappa}\int f(R,T)\sqrt{-g}~d^4x+\int L_{m}\sqrt{-g}~d^4x,
\end{eqnarray}
Here the variable of the Lagrangian functional is the inverse of the metric tensor $g^{\mu\nu}$. Assume that there is no the derivatives of $g^{\mu\nu}$ in $T_{\mu\nu}$. $L_{m}$ is the Lagrangian of the matter field and the corresponding energy-momentum tensor is
\begin{eqnarray}\label{1.2}
 \nonumber   T_{\mu\nu}&:=&-\frac{2}{\sqrt{-g}}\frac{\delta(\sqrt{-g}L_{m})}{\delta g^{\mu\nu}}\\
&=&g_{\mu\nu}L_{m}-2\frac{\partial L_{m}}{\partial g^{\mu\nu}}.
\end{eqnarray}
It follows the variation of $T_{\mu\nu}$ which is
\begin{eqnarray}\label{1.2a}
    \nonumber\delta T_{\mu\nu}&=&L_{m}\delta g_{\mu\nu}+g_{\mu\nu}\frac{\partial L_{m}}{\partial g^{\alpha\beta}}\delta g^{\alpha\beta}-2\frac{\partial^2 L_{m}}{\partial g^{\mu\nu}\partial g^{\alpha\beta}}\delta g^{\alpha\beta}\\
\nonumber&=&L_{m}\delta g_{\mu\nu}+g_{\mu\nu}\frac{\partial L_{m}}{\partial g^{\alpha\beta}}\delta g^{\alpha\beta}-2\frac{\partial^2 L_{m}}{\partial g^{\mu\nu}\partial g^{\alpha\beta}}\delta g^{\alpha\beta}\\
\nonumber&=&-L_{m}g_{\mu\alpha}g_{\nu\beta}\delta g^{\alpha\beta}+\frac{g_{\mu\nu}}{2}(g_{\alpha\beta}L_{m}-T_{\alpha\beta})\delta g^{\alpha\beta}\\
&-&2\frac{\partial^2 L_{m}}{\partial g^{\mu\nu}\partial g^{\alpha\beta}}\delta g^{\alpha\beta}.
\end{eqnarray}
Hence one will obtain
\begin{eqnarray}\label{1.2a}
  \nonumber \frac{\delta T_{\mu\nu}}{\delta g^{\alpha\beta}}&=&-L_{m}g_{\mu\alpha}g_{\nu\beta}+\frac{1}{2}g_{\mu\nu}g_{\alpha\beta}L_{m}\\
&-&\frac{1}{2}g_{\mu\nu}T_{\alpha\beta}-2\frac{\partial^2 L_{m}}{\partial g^{\mu\nu}\partial g^{\alpha\beta}}.
\end{eqnarray}

An difference from the standard Einstein-Hilbert variation is that we need to deal with the variation of the trace of the energy-momentum tensor. We could define a new object $\Theta_{\mu\nu}$ by
\begin{eqnarray}\label{1.3}
    \frac{\delta~T}{\delta g^{\mu\nu}}=T_{\mu\nu}+g^{\alpha\beta}\frac{\delta~T_{\alpha\beta}}{\delta g^{\mu\nu}}
    =T_{\mu\nu}+\Theta_{\mu\nu}.
\end{eqnarray}
By means of (\ref{1.2a}), one can reexpress
\begin{eqnarray}\label{1.3a}
   \Theta_{\mu\nu}=-L_{m}g_{\mu\nu}+2g_{\mu\nu}L_{m}
-2T_{\mu\nu}-2g^{\alpha\beta}\frac{\partial^2 L_{m}}{\partial g^{\mu\nu}\partial g^{\alpha\beta}}.
\end{eqnarray}

One will obtain the field equation after doing variation
\begin{eqnarray}\label{1.4}
 \nonumber   &&f_{R}R_{\mu\nu}+\nabla^{\alpha}\nabla_{\alpha}(f_{R}g_{\mu\nu})-\nabla_{\mu}\nabla_{\nu}f_{R}\\
    &&+f_{T}T_{\mu\nu}+f_{T}\Theta_{\mu\nu}
    -\frac{1}{2}g_{\mu\nu}f=kT_{\mu\nu},
\end{eqnarray}
where and elsewhere we will denote the abbreviation to express the partial derivative, such as $f_{R}:=\frac{\partial f(R,T)}{\partial R}$, $f_{T}:=\frac{\partial f(R,T)}{\partial T}$ and so on.
To contract the field equation will result in
\begin{eqnarray}\label{1.5}
    f_{R}R+3\nabla^{\alpha}\nabla_{\alpha}f_{R}-2f=kT-f_{T}T-f_{T}\Theta
\end{eqnarray}

In order to observe explicitly the behavior of the field equations, we will pay our attention to the specific matter field -- the perfect fluid which is expressed by
\begin{eqnarray}\label{1.6}
    T_{\mu\nu}=(\rho+p)U_{\mu}U_{\nu}+pg_{\mu\nu}
\end{eqnarray}
Note that for a scalar field, $L_{m}=p$, thus from (\ref{1.3a}) and (\ref{1.6})
\begin{eqnarray}\label{1.7}
    \Theta_{\mu\nu}=-2T_{\mu\nu}+pg_{\mu\nu}
\end{eqnarray}
Insert (\ref{1.7}) into (\ref{1.4}) and the field equation will be reduced to be
\begin{eqnarray}\label{1.8}
 \nonumber &&f_{R}R_{\mu\nu}+g_{\mu\nu}\nabla^{\alpha}\nabla_{\alpha}f_{R}-\nabla_{\mu}\nabla_{\nu}f_{R}\\
  &&=(k+f_{T})T_{\mu\nu}-f_{T}pg_{\mu\nu}+\frac{f(R,T)}{2}g_{\mu\nu}
\end{eqnarray}

\subsection{The case $f(R,T)=R+F(T)$}

In the following, let's focus on the specific case $f(R,T)=R+F(T)$ with $T=-\rho+3p$. In that case, we will obtain from (\ref{1.8})
the generalized field equation which read
\begin{eqnarray}\label{1.11}
    R_{\mu\nu}-\frac{1}{2}g_{\mu\nu}R
    =(\kappa+F_{T})T_{\mu\nu}-pF_{T}g_{\mu\nu}+\frac{F(T)}{2}g_{\mu\nu}
\end{eqnarray}
We could regard the $f(R,T)$ gravity as a model with the effective gravitational constant $G_{eff}$ and the effective cosmological constant $\Lambda_{eff}$. In that way, we could define
\begin{eqnarray}\label{1.12}
G_{eff}:=\kappa+F_{T},~~\Lambda_{eff}:=pF_{T}-\frac{F(T)}{2}.
\end{eqnarray}
Obviously, both $G_{eff}$ and $\Lambda_{eff}$ are time-varying which reflects the characteristic interaction between the matter and geometry in this model.

Although (\ref{1.11}) is more complicated than the standard Einstein equation, it is still legal to employ the Bianchi identity to obtain some kind of ``conservation''. Hence, based on the Bianchi identity, we will obtain
\begin{eqnarray}\label{1.13}
\nonumber\nabla^{\mu}G_{\mu\nu} &=&(\kappa+F_{T})\nabla^{\mu}T_{\mu\nu}+ (T_{\mu\nu}-pg_{\mu\nu})\nabla^{\mu}F_{T}\\
&+&  \frac{1}{2}g_{\mu\nu}F_{T}\nabla^{\mu}p-\frac{1}{2}g_{\mu\nu}F_{T}\nabla^{\mu}\rho\equiv0
\end{eqnarray}

\section{The relation between equation of state and $F(T)$}

\subsection{FLRW equations in the model}

In the isotropic and homogeneous cosmology, the metric is
\begin{eqnarray}\label{2.1}
  ds^2=-dt^2+a(t)^2[\frac{dr^2}{1-Kr^2}+r^2(d\theta^2+\sin^2\theta d\varphi^2)]
\end{eqnarray}
with $K=+1, 0, -1$ respectively to represent the three kinds of three-geometry of constant curvature.

Based on the metric (\ref{2.1}), the two field equations (FLRW equations) are
\begin{eqnarray}\label{2.2.1}
  3(\frac{\dot{a}^2}{a^2}+\frac{K}{a^2})&=&(\kappa+F_{T})\rho+pF_{T}-\frac{1}{2}F, \\
  \label{2.2.2} -(2\frac{\ddot{a}}{a}+\frac{\dot{a}^2}{a^2}+\frac{K}{a^2})&=&\kappa p+\frac{1}{2}F.
\end{eqnarray}
Another useful form, to combine (\ref{2.2.1}) and (\ref{2.2.2}), is
\begin{eqnarray}\label{2.2.3}
 -6\frac{\ddot{a}}{a}=(\kappa+F_{T})\rho+(3\kappa+F_{T}) p+F.
\end{eqnarray}

There are two unknowns in these equations; one is the EOS, namely, the exact relation between $\rho$ and $p$, the other is the function $F(T)$. However, since $F(T)$ is the function of $\rho$ and $p$, it must be closely tied to the EOS. In view of this, we will adopt the adiabatic process to fix the pattern of $F(T)$ and then when the EOS is obtained we are able to solve the field equations (\ref{2.2.1}, \ref{2.2.2}) definitely.



\subsection{Classification of $F(T)$}

First, let $\nu=0$ and (\ref{1.13}) shows
\begin{eqnarray}\label{2.3}
\nonumber&&(\kappa+F_{T})[-\dot{\rho}-3\frac{\dot{a}}{a}(\rho+p)]\\
&&-(\rho+p)\frac{dF_{T}}{dt}+ \frac{1}{2}F_{T}(\dot{p}-\dot{\rho})=0,
\end{eqnarray}
which also can be rewritten in the following way,
\begin{eqnarray}\label{2.4}
\nonumber&&\frac{d}{dt}[a^3(\rho+p)(\kappa+F_{T})]\\
&&-a^3(\kappa+\frac{3}{2}F_{T})\dot{p}+\frac{a^3}{2}F_{T}\dot{\rho}=0.
\end{eqnarray}

When applying the thermodynamics into cosmology, one often invokes the process of adiabatic expansion to simplify the model. Thus, we will employ this assumption as the first step to analysis our equations.
The thermodynamic first law states
\begin{eqnarray}\label{3.1}
\nonumber dU+pdV&=&d(\rho a^3)+pda^3\\
&=&d\rho+3(\rho+p)da^3=TdS.
\end{eqnarray}
The adiabatic expansion is compatible with the conservation of entropy, which means $dS=0$ and results in
\begin{eqnarray}\label{3.2}
\dot{\rho}+3H(\rho+p)=0.
\end{eqnarray}

Assume that the EOS satisfies the relation $p=w\rho$ with $w$ a constant. We are readily from (\ref{3.2}) to obtain
\begin{eqnarray}\label{3.4}
\rho=\rho_{0}a_{0}^{3(w+1)}a^{-3(w+1)},
\end{eqnarray}
where $\rho_{0}$ and $a_{0}$ are the present values of the density of matter field and the scale factor.
The value of $\rho_{0}$ is always regarded to be positive in this paper.

By the way, introduce two abbreviations $\rho_{*}$ and $A$ and denote
\begin{eqnarray}\label{3.4a}
\nonumber\rho&=&\rho_{0}a_{0}^{3(w+1)}a^{-3(w+1)}\\
&:=&\rho_{*}^{3(w+1)}a^{-3(w+1)}=A^{-3(w+1)},
\end{eqnarray}
where $A$ can be considered as ``reduced`` scale factor for convenience.

To combine (\ref{2.3}) and (\ref{3.2}), it's readily to get
\begin{eqnarray}\label{3.5}
-(1+w)\rho\frac{dF_{T}}{dt}+  \frac{1}{2}F_{T}(w-1)\dot{\rho}=0.
\end{eqnarray}
The equation (\ref{3.5}) apparently displays that the exact form of $F(T)$ depends on the corresponding parameter $w$. Consequently, let's deduce the exact pattern of $F(T)$ for the different parameter $w$ in the following. According to (\ref{3.5}), we could classify these $F(T)$ in terms of different $w$.

(a) If $w=-1$, $F_{T}\dot{\rho}=0$. Besides, the equation (\ref{3.2}) tells that $\rho$ is constant. In this case, (\ref{3.5}) has no restriction on $F_{T}$.

(b) If $w=1$, it turns out $F_{T}=c_{1}$ a constant so that $F(T)=c_{1}T+c_{2}$, namely $F(\rho)=2c_{1}+c_{2}$.

(c) If $w=\frac{1}{3}$, that $T=-\rho+3p\equiv0$ will result $F_{T}=0$, namely $F(T)=c_{2}$ a constant.

(d) If $w\neq\pm1, \frac{1}{3}$, the equation (\ref{3.5}) will provide
\begin{eqnarray}\label{3.6}
F_{T}=c_{1}\rho^{\frac{w-1}{2(w+1)}},
\end{eqnarray}
with $\alpha$ a constant. Since $F_{\rho}=(3w-1)F_{T}$, we have
\begin{eqnarray}\label{3.7}
F_{\rho}=c_{1}(3w-1)\rho^{\frac{w-1}{2(w+1)}}.
\end{eqnarray}
Note that the class (d) in fact contains two subclasses.

(d1) If $\frac{w-1}{2(w+1)}=-1$ or equivalently $w=-\frac{1}{3}$, then
\begin{eqnarray}\label{3.8}
F(\rho)=-2c_{1}\ln\rho+c_{2},
\end{eqnarray}
where $c_{1}$, $c_{2}$ are constants.

(d2) If $\frac{w-1}{2(w+1)}\neq-1$ or equivalently $w\neq-\frac{1}{3}$, then
\begin{eqnarray}\label{3.9}
F(\rho)=2c_{1}\frac{(3w-1)(w+1)}{3w+1}\rho^{\frac{3w+1}{2(w+1)}}+c_{2},
\end{eqnarray}
where $c_{1}$, $c_{2}$ are constants.

\section{Evolution of the scale factor}

We will study the behavior of the scale factor by employing the five kinds of $F(T)$ which are corresponding to $w=\pm1, \pm=\frac{1}{3}, 0$ respectively. Different $F(T)$ based on different $w$ and different parameters $c_{1}$ and $c_{2}$ will result in the specific mode of the scale factor. We conclude that these modes can be classified into nine types. Table \ref{tab1} shows simply the characters for each type and the schematic diagrams of different types of universe are displayed in this section.

Here we will list all modes of the scale factor $A(t)$\footnote{In the following, we will still call $A(t)$ scale factor rather than ``reduced scale factor`` for convenience.} with regard to several specific equations of state $w=\frac{p}{\rho}$. For the sake of brevity, we will assume that $w$ is constant and focus on the five categories, $w=\pm1, \pm\frac{1}{3}, 0$. Every mode coincides with its particular range of scale factor, which closely relates to how the corresponding universe evolves.

In the following, we will examine flat 3-space, i.e., $K=0$ and write down the concrete form of scale factor for every mode if they can be expressed in terms of elementary function.

\begin{table}[tbp]\tiny
\centering
\begin{tabular}{|c|c|c|c|c|}
  \hline
   type & &initial state  &  final state & \textrm{feature} \\  \hline

1 & close &$A(t_{i})=0$  & $A(t_{f})=0$& $\dot{A}>0,\ddot{A}<0\rightarrow\dot{A}<0,\ddot{A}<0$\\
2 & open & $A(t_{i})=0$ & $A(\infty)=\infty$& $\dot{A}>0,\ddot{A}<0\rightarrow\dot{A}>0,\ddot{A}>0$\\
3 & open & $A(t_{i})>0$ & $A(\infty)=\infty$ & $\dot{A}>0,\ddot{A}>0$\\
4 & bound & $A(t_{i})=0$ & $A(\infty)=A_{f}$& $\dot{A}>0,\ddot{A}<0$\\
5 & static & $A(t_{i})>0$ & $A(\infty)=A_{i}$& $A$ unchanged\\
6 & open & $A(-\infty)=0$ & $A(\infty)=\infty$& $\dot{A}>0,\ddot{A}>0$\\
7 & open & $A(t_{i})=0$ & $A(\infty)=\infty$& $\dot{A}>0,\ddot{A}<0$\\
8 & open & $A(t_{i})=0$ & $A(\infty)=\infty$& $\dot{A}>0,\ddot{A}>0$\\
9 & close & $A(t_{i})=0$ & $A(t_{f})=0$ & $\dot{A}>0,\ddot{A}>0\rightarrow\dot{A}>0,\ddot{A}<0\rightarrow\dot{A}<0,\ddot{A}<0\rightarrow\dot{A}<0,\ddot{A}>0$\\
\hline

\end{tabular}
\caption{\label{tab1} Type of the universe }
\end{table}


\begin{figure}
  \begin{minipage}[tbp]{0.5\linewidth}
    \centering
    \includegraphics[trim=5 5 4.65 5, clip,height=8cm, width=6cm,angle=270]{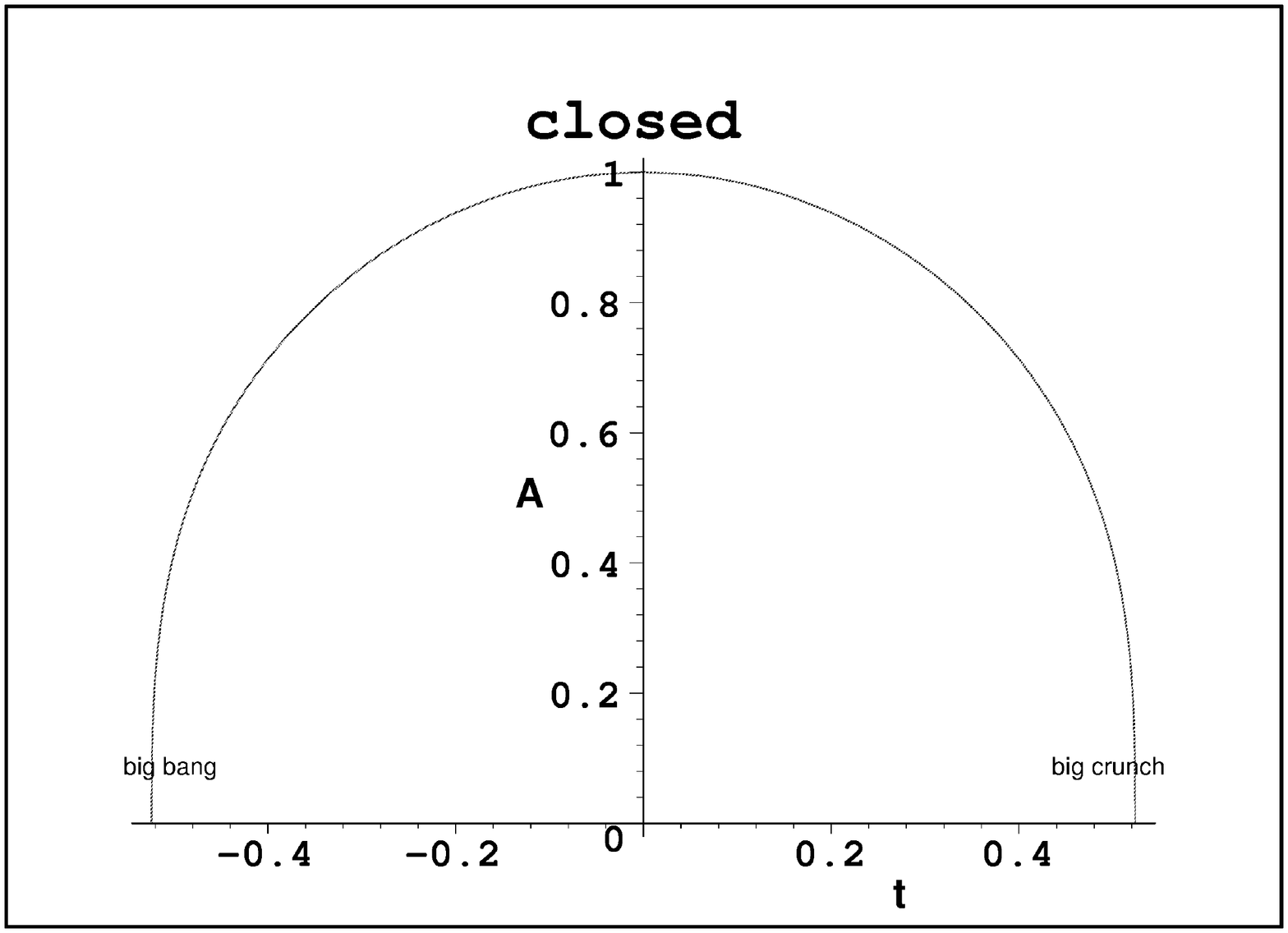}
    \caption{\scriptsize \textbf{Type 1}: Closed with inflection point.
    \protect \\ Choose the mode (2-6) with $\kappa+c_{1}=3,~c_{2}=6$.}
    \label{fig1}
  \end{minipage}
  \begin{minipage}[tbp]{0.5\linewidth}
    \centering
    \includegraphics[trim=5 5 4.65 5, clip,height=8cm, width=6cm,angle=270]{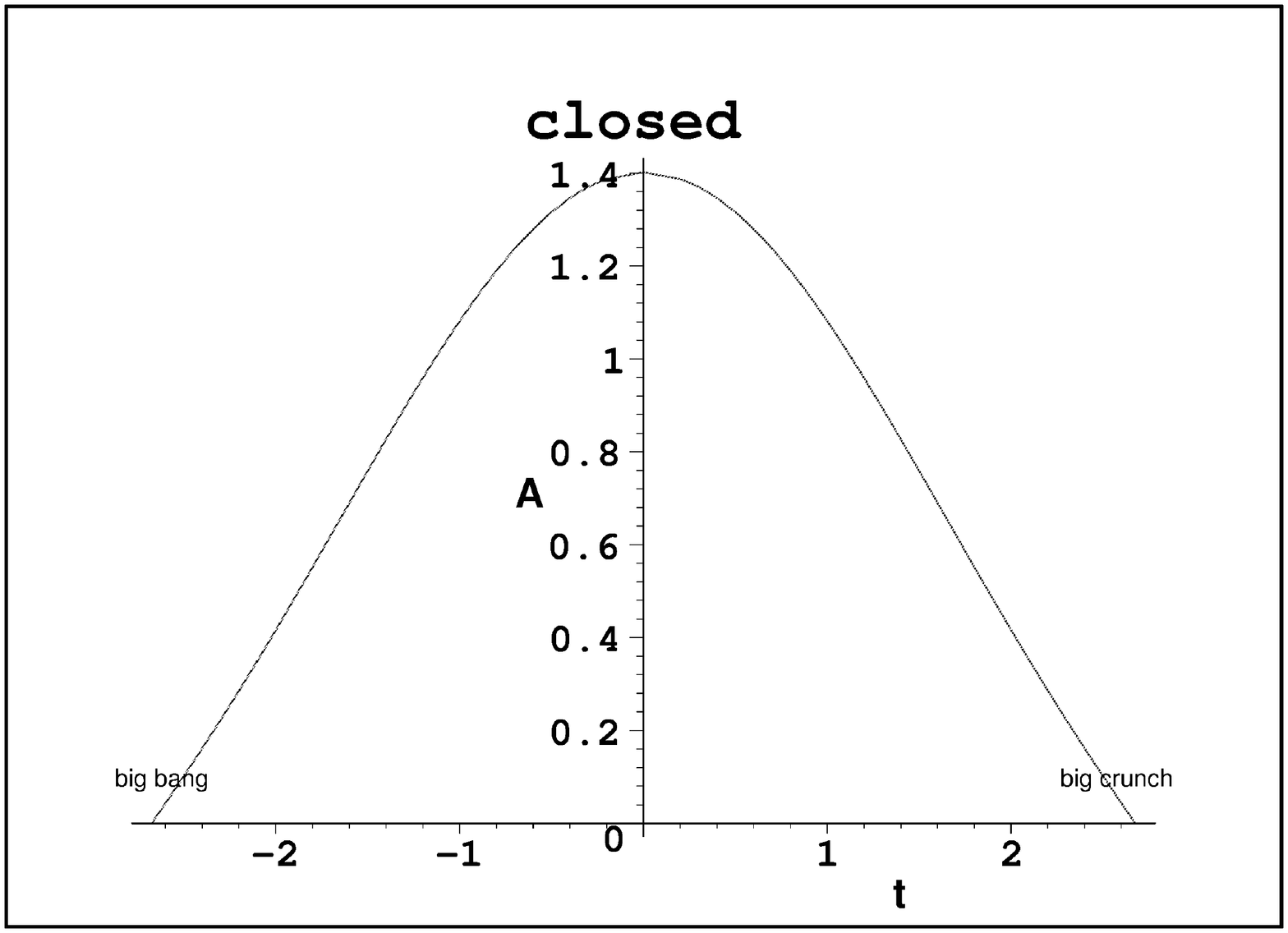}
    \caption{\scriptsize
    \label{fig2} \textbf{Type 9}: Closed with two inflection point.
    \protect \\ Choose the mode (4-3) with $\alpha=-2$, $\beta=\frac{1}{6}$, $\gamma=1$.}
  \end{minipage}
   \begin{minipage}[tbp]{0.5\linewidth}
    \centering
    \includegraphics[trim=5 5 4.65 5, clip,height=8cm, width=6cm,angle=270]{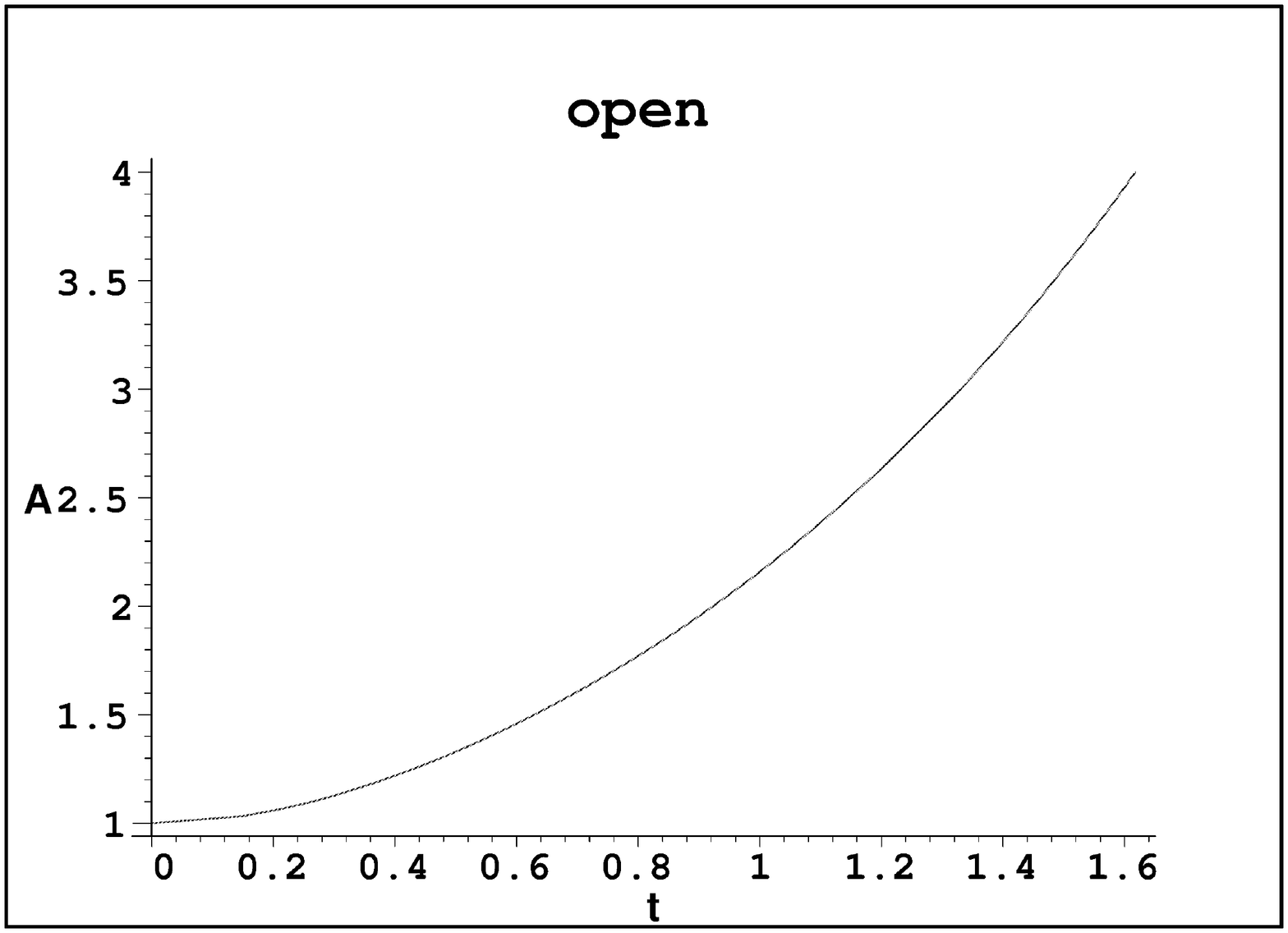}
    \caption{\scriptsize \textbf{Type 3}: Expand forever from a finite scale.
    \protect \\ Choose the mode (2-8) with $\kappa+c_{1}=-3$, $c_{2}=-6$.}
    \label{fig3}
  \end{minipage}
   \begin{minipage}[tbp]{0.5\linewidth}
    \centering
    \includegraphics[trim=5 5 4.65 5, clip,height=8cm, width=6cm,angle=270]{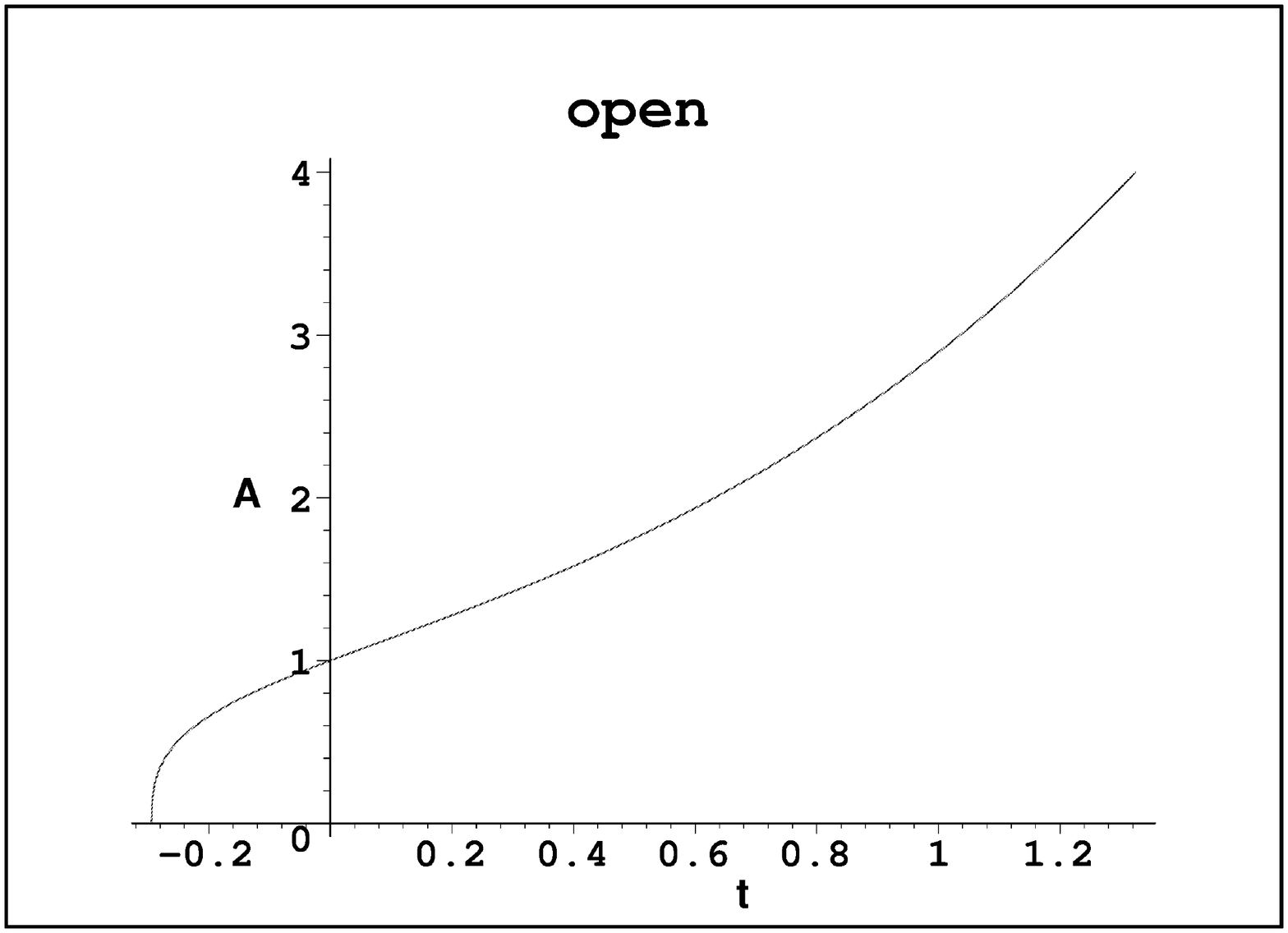}
    \caption{\scriptsize
    \label{fig4} \textbf{Type 2}: Expand forever with an inflection.
    \protect \\ Choose the mode (2-7) with $\kappa+c_{1}=3$, $c_{2}=-6$.}
  \end{minipage}
   \begin{minipage}[tbp]{0.5\linewidth}
    \centering
    \includegraphics[trim=5 5 4.65 5, clip,height=8cm, width=6cm,angle=270]{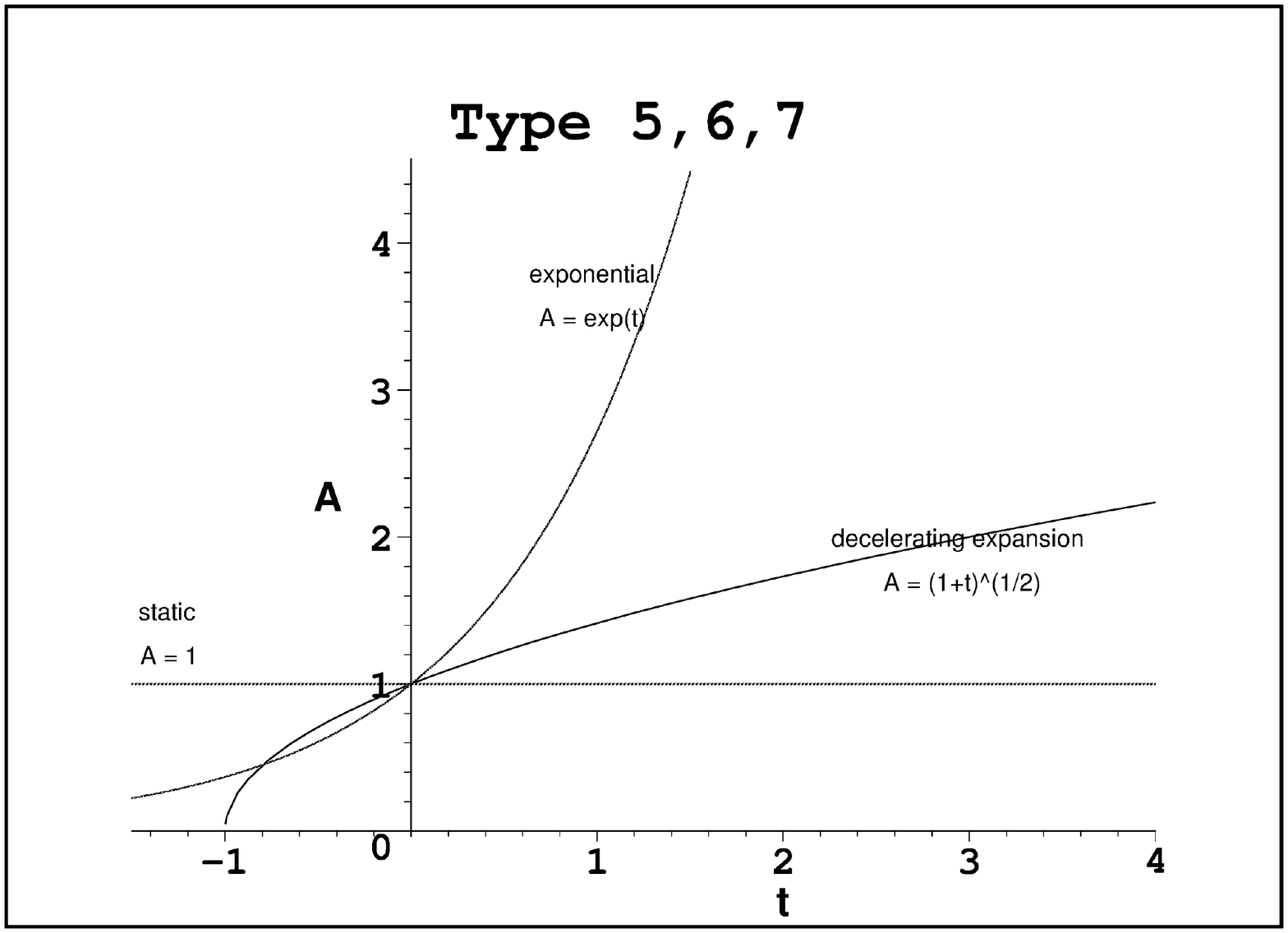}
    \caption{\scriptsize  Given $A_{0}=1$.
    \protect\\\textbf{Type 5:} Static.
     \protect\\\textbf{Type 6:} Exponential. Choose $c_{2}=-6$.
     \protect \\\textbf{Type 7:} Expand at ever-decreasing rate.
     \protect \\ Choose $\kappa+c_{1}=\frac{1}{3}$. }
    \label{fig5}
  \end{minipage}%
  \begin{minipage}[tbp]{0.5\linewidth}
    \centering
    \includegraphics[trim=5 5 4.65 5, clip,height=8cm, width=6cm,angle=270]{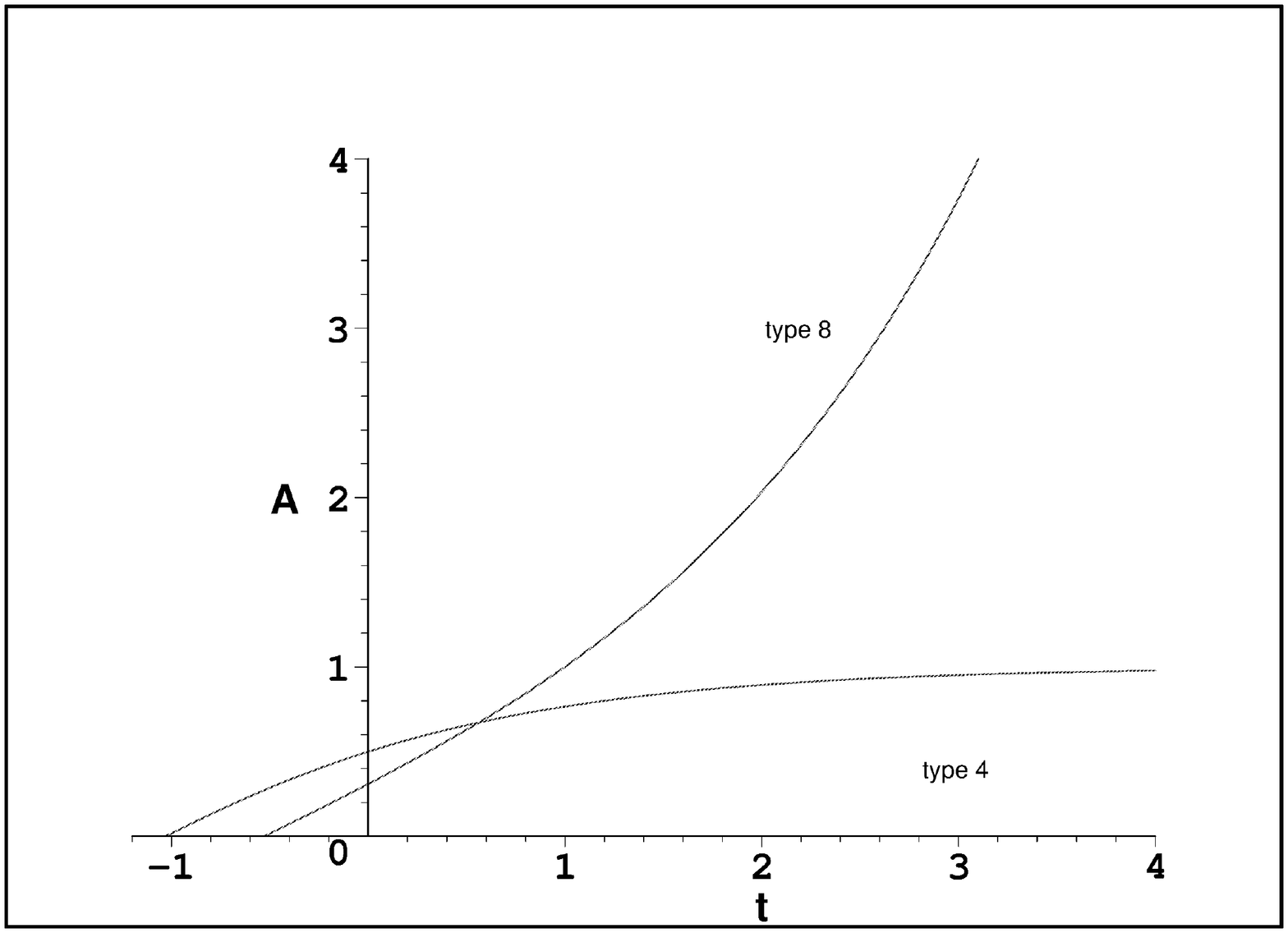}
    \caption{\scriptsize \textbf{Type 4}: Upper bound. Expand at an ever-decreasing rate.
    \protect \\Choose mode the (4-5) with $\alpha=2$, $\beta=-1$, $\gamma=1$.
    \protect \\ \textbf{Type 8}: Expand at an ever-increasing rate.
    \protect \\ Choose the mode (4-2) with $c_{2}=-2$. }
    \label{fig6}
  \end{minipage}
\end{figure}


\subsection{category 1: $w=-1$}

Firstly, add (\ref{2.2.1}) and (\ref{2.2.2}) together to obtain
\begin{eqnarray}\label{a.1.1}
\frac{\ddot{A}}{A}-\frac{\dot{A}^2}{A^2}-\frac{K}{\rho_{*}^2A^2}=0.
\end{eqnarray}
Multiply $H$ to the above equation and integrate it to obtain
\begin{eqnarray}\label{a.1.2}
 H^2+\frac{K}{\rho_{*}^2A^2}=\lambda
\end{eqnarray}
with $\lambda$ a integration constant.
To compare (\ref{a.1.2}) with (\ref{2.2.1}), we will know that $\kappa\rho-\frac{F}{2}$ must be a constant. Since $\rho$ is constant for $w=-1$, it means that $F$ should be a constant here.

When $K=0$, $\lambda\geq0$. One can choose some specific value $H_{*}^2$ to be $\lambda$. Integrate (\ref{a.1.2}) to get
$A(t)=e^{H_{*}(t+t_{*})}$ with an integration constant $t_{*}$. Given $A=A_{0}$ as $t=0$, one will have
$t_{*}=\frac{\ln{A_{0}}}{H_{*}}$ and
\begin{eqnarray}
A(t)=A_{0}e^{H_{*}t}.
\end{eqnarray}
Clearly, this solution describes a monotonic scale factor $A(t)$. Besides, (\ref{1.12}) indicates
\begin{eqnarray}
G_{eff}=\kappa,~\Lambda_{eff}=-\frac{c_{2}}{2}.
\end{eqnarray}

\subsection{category 2: $w=1$}
Follow (\ref{2.2.1}) and (\ref{3.4a}) to get
\begin{eqnarray}\label{a.2.1}
\pm(t+t_{*})=\int\frac{\sqrt{3}A^2 dA}{\sqrt{\kappa+c_{1}-\frac{c_{2}}{2}A^6}},
\end{eqnarray}
with an integration constant $t_{*}$. The necessary condition of the allowable $A$ is $A\geq0$.
Furthermore, the condition $\kappa+c_{1}-\frac{c_{2}}{2}A^6\geq0$ also needs to be satisfied.
The scale factor of this category can be classified into 6 modes and can be expressed by elementary functions. They are listed in Table \ref{tab2}. Here we will examine
the relation between
$A$ and its 2nd derivative $\ddot{A}$.

If there exists some $A_{I}$ satisfying $\ddot{A}_{I}=0$, according to (\ref{2.2.2}) and (\ref{3.8}),
\begin{eqnarray}\label{a.2.2}
A_{I}^6=\frac{-\kappa}{H_{I}^2+\frac{c_{2}}{2}},
\end{eqnarray}
where $H_{I}$ is the value of Hubble constant at $A_{I}$. Apparently, it implies that $c_{2}<\frac{-H_{I}^2}{2}\leq0$
is a necessary condition once $\ddot{A}_{I}=0$ occurs. In that case, only the four modes (2-2), (2-4), (2-7), (2-8)  in which the scale factor will concave upward early and concave downward later or vice versa could
be saved. The mode (2-4) can be discarded
right away since it is an exponential function which is always lack of $A_{I}$
no matter the value of $c_{2}$ is.

Let's examine the mode (2-7). Based on (\ref{2.2.3}), if there exists some $A_{I}$, then
\begin{eqnarray}\label{a.2.3}
A_{I}^6=\frac{-4(\kappa+c_{1})}{c_{2}}=\frac{4(\kappa+c_{1})}{|c_{2}|}>0.
\end{eqnarray}
The allowable range $A$ of this mode lies in the whole non-negative interval such that $A_{I}$ is bound to be in existence.
In addition, (\ref{2.2.3}) implies that $\ddot{A}$ must be positive for this mode as $A$ is large enough.
Thus, the mode (2-7) belongs to the type 2. Similarly, the mode (2-2) is also classified into the type 2.

Turn to the mode (2-8). Notice that
\begin{eqnarray}\label{a.2.4}
A\geq (\frac{2|\kappa+c_{1}|}{|c_{2}|})^{\frac{1}{6}}>0>-\frac{4|\kappa+c_{1}|}{|c_{2}|},
\end{eqnarray}
which implies that $A_{I}$ is in the absence for the mode (2-8). Due to its growing 2nd derivative $\ddot{A}$,
like the mode (2-7), one could realize that this mode belongs to the type 3.

\begin{table}[tbp]\tiny
\centering
\begin{tabular}{|c|c|c|c|}
  \hline
  $w=1$ & \textrm{mode} &  \textrm{reduced scale factor} $A$ & \textrm{type} \\ \hline

  (2-1) &$c_{1}=0,~c_{2}>0$ & $A=[\sqrt{\frac{2\kappa}{c_{2}}-A^6_{0}}\sin\sqrt{\frac{3c_{2}}{2}}t+A^3_{0}\cos\sqrt{\frac{3c_{2}}{2}}t~]^{\frac{1}{3}},~0\leq A\leq (\frac{2\kappa}{c_{2}})^{\frac{1}{6}}$ & \textbf{\textrm{1}}\\

  (2-2) &$c_{1}=0,~c_{2}<0$ &$A=[\sqrt{\frac{2\kappa}{|c_{2}|}+A^6_{0}}\sinh\sqrt{\frac{3|c_{2}|}{2}}t+A^3_{0}\cosh\sqrt{\frac{3|c_{2}|}{2}}t~]^{\frac{1}{3}},~ A\geq0$ &\textbf{\textrm{2}} \\

  (2-3) &$c_{1}\neq0,~\kappa+c_{1}=0,~c_{2}=0$ & $A=A_{0}$ &\textbf{\textrm{5}} \\

  (2-4) &$c_{1}\neq0,~\kappa+c_{1}=0,~c_{2}<0$ & $A=A_{0}\exp{(\sqrt{\frac{|c_{2}|}{6}}t)},~A\geq0$  &\textbf{\textrm{6}} \\

  (2-5) &$c_{1}\neq0,~\kappa+c_{1}>0,~c_{2}=0$ & $A=[3(\kappa+c_{1})~t+A^3_{0}~]^{\frac{1}{3}},~A\geq0$ &\textbf{\textrm{7}} \\

  (2-6) &$c_{1}\neq0,~\kappa+c_{1}>0,~c_{2}>0$ & $A=[\sqrt{\frac{2(\kappa+c_{1})}{c_{2}}-A^6_{0}}\sin\sqrt{\frac{3c_{2}}{2}}t+A^3_{0}\cos\sqrt{\frac{3c_{2}}{2}}t~]^{\frac{1}{3}},~0\leq A\leq (\frac{2(\kappa+c_{1})}{c_{2}})^{\frac{1}{6}}$ &\textbf{\textrm{1}} \\

  (2-7) &$c_{1}\neq0,~\kappa+c_{1}>0,~c_{2}<0$ & $A=[\sqrt{\frac{2(\kappa+c_{1})}{|c_{2}|}+A^6_{0}}\sinh\sqrt{\frac{3|c_{2}|}{2}}t+A^3_{0}\cosh\sqrt{\frac{3|c_{2}|}{2}}t~]^{\frac{1}{3}},~ A\geq0$ &\textbf{\textrm{2}} \\

  (2-8) &$c_{1}\neq0,~\kappa+c_{1}<0,~c_{2}<0$ & $A=[\sqrt{-\frac{2(|\kappa+c_{1}|)}{|c_{2}|}+A^6_{0}}\sinh\sqrt{\frac{3|c_{2}|}{2}}t+A^3_{0}\cosh\sqrt{\frac{3|c_{2}|}{2}}t~]^{\frac{1}{3}},~ A\geq(\frac{2(|\kappa+c_{1}|)}{|c_{2}|})^{\frac{1}{6}}$ &\textbf{\textrm{3}} \\
  \hline
\end{tabular}
\caption{\label{tab2} Given the scale factor $A$ equal to $A_{0}$ as $t=0$.}
\end{table}
Note that in principle the modes (2-1) and (2-2) are the same as (2-6) and (2-7) respectively.

 For the category with $w=1$, (\ref{1.12}) indicates
\begin{eqnarray}
G_{eff}=\kappa+c_{1},~\Lambda_{eff}=-\frac{c_{2}}{2}.
\end{eqnarray}


\subsection{category 3: $w=\frac{1}{3}$}
Combine (\ref{2.2.1}) with (\ref{3.4a}) to get
\begin{eqnarray}\label{a.3.1}
\pm(t+t_{*})=\int\frac{\sqrt{3}AdA}{\sqrt{\kappa-\frac{c_{2}}{2}A^4}},
\end{eqnarray}
with the integration constant $t_{*}$.
The scale factor $A$ satisfies $A\geq0$ and the condition $\kappa-\frac{c_{2}}{2}A^4\geq0$.
There are two modes here, mode (3-1) characterized by $c_{2}>0$ and (3-2) characterized by $c_{2}<0$. Both of them will yield the scalar factor expressed in terms of elementary functions that are exhibited in Table \ref{tab3}.

As we did before, here examine the behavior of 2nd derivative for the two modes. To follow (\ref{2.2.3}),
if there exists some $A_{I}$ satisfying $\ddot{A}_{I}=0$,
\begin{eqnarray}
A_{I}^4=\frac{-\kappa}{c_{2}}.
\end{eqnarray}
Obviously, the mode (3-1) contains no inflection point $A_{I}$ and results in a totally concave downward shape, the type 1.
Conversely, the mode (3-2) whose 2nd derivative is approaching to infinity as $A$ is running far away from one
is afford to contain an $A_{I}$. Thus, the mode (3-2) belongs to the type 2.

On the other hand, (\ref{1.12}) states
\begin{eqnarray}
G_{eff}=\kappa,~\Lambda_{eff}=-\frac{c_{2}}{2}.
\end{eqnarray}

\begin{table}[tbp]\tiny
\centering
\begin{tabular}{|c|c|c|c|}
  \hline
 $w=\frac{1}{3}$ & \textrm{mode}  &  \textrm{reduced scale factor} $A$ & \textrm{type} \\ \hline
  (3-1) &$c_{2}>0$ & $A=[\sqrt{\frac{2\kappa}{c_{2}}-A^4_{0}}\sin\sqrt{\frac{2c_{2}}{3}}t+A^2_{0}\cos\sqrt{\frac{2c_{2}}{3}}t~]^{\frac{1}{2}},~0\leq A\leq (\frac{2\kappa}{c_{2}})^{\frac{1}{4}}$  &\textbf{\textrm{1}} \\
  (3-2) &$c_{2}<0$&
    $A=[\sqrt{\frac{2\kappa}{|c_{2}|}+A^4_{0}}\sinh\sqrt{\frac{2|c_{2}|}{3}}t+A^2_{0}\cosh\sqrt{\frac{2|c_{2}|}{3}}t~]^{\frac{1}{2}},~ A\geq0$
&\textbf{\textrm{2}} \\
\hline
\end{tabular}
\caption{\label{tab3} Given the scale factor $A$ equal to $A_{0}$ as $t=0$. }
\end{table}


\subsection{category 4: $w=-\frac{1}{3}$}
Combine (\ref{2.2.1}) with (\ref{3.4a}) to get
\begin{eqnarray}\label{a.4.1}
\pm(t+t_{*})=\int\frac{\sqrt{3}dA}{\sqrt{\kappa+(\frac{2}{3}c_{1}-\frac{c_{2}}{2})A^2-2c_{1}A^2\ln{A}}},
\end{eqnarray}
with the integration constant $t_{*}$. The scale factor $A$ satisfies two conditions,
$A\geq0$ and $\gamma+\beta A^2+\alpha A^2\ln{A}\geq0$.
Here denote $\alpha=-2c_{1}$, $\beta=\frac{2}{3}c_{1}-\frac{c_{2}}{2}$ and $\gamma=\kappa$.

\begin{table}[tbp]\tiny
\centering
\begin{tabular}{|c|c|c|c|}
  \hline
    $w=-\frac{1}{3}$ & \textrm{mode}  &  \textrm{reduced scale factor} $A$ & \textrm{type} \\ \hline
  (4-1) &$c_{1}=0,~ c_{2}>0$ & $A=\sqrt{\frac{2\kappa}{c_{2}}-A^2_{0}}\sin\sqrt{\frac{c_{2}}{6}}t+A_{0}\cos\sqrt{\frac{c_{2}}{6}}t,~0\leq A\leq (\frac{2\kappa}{c_{2}})^{\frac{1}{2}}$  &\textbf{\textrm{1}} \\

(4-2) &$c_{1}=0,~ c_{2}<0$ & $A=\sqrt{\frac{2\kappa}{|c_{2}|}+A^2_{0}}\sinh\sqrt{\frac{|c_{2}|}{6}}t+A_{0}\cosh\sqrt{\frac{|c_{2}|}{6}}t,~ A\geq0$&\textbf{\textrm{8}} \\

$\begin{array}{l}\textrm{(4-3)}\\
\\
\\
\end{array}$ & $\alpha<0$ & $\begin{array}{c} \textrm{not elementary function}\\
0\leq A\leq\exp{[\frac{1}{2}\textrm{W}_{0}(\frac{-2\gamma}{\alpha}e^{\frac{2\beta}{\alpha}})-\frac{\beta}{\alpha}]}\end{array}$&  \textbf{\textrm{9}} \\

  $\begin{array}{l}\textrm{(4-4)}\\
\\
\\
\end{array}$
 & $\begin{array}{l}
\alpha>0,~e^{-1}>\frac{2\gamma}{\alpha}e^{\frac{2\beta}{\alpha}}\\
\end{array}$ & $\begin{array}{c} \textrm{not elementary function}\\
0\leq A\leq\exp{[\frac{1}{2}\textrm{W}_{-1}(\frac{-2\gamma}{\alpha}e^{\frac{2\beta}{\alpha}})-\frac{\beta}{\alpha}]}~
\textrm{or}~A\geq\exp{[\frac{1}{2}\textrm{W}_{0}(\frac{-2\gamma}{\alpha}e^{\frac{2\beta}{\alpha}})-\frac{\beta}{\alpha}]}
\end{array}$&  \textbf{\textrm{1 or 3}}\\

    $\begin{array}{l}\textrm{(4-5)}\\
 \\
\end{array}$& $\alpha>0,~e^{-1}=\frac{2\gamma}{\alpha}e^{\frac{2\beta}{\alpha}}$ & $\begin{array}{c} \textrm{not elementary function} \\
A\geq\exp{[-\frac{1}{2}-\frac{\beta}{\alpha}]}
~\textrm{or}~0\leq A\leq\exp{[-\frac{1}{2}-\frac{\beta}{\alpha}]}\end{array}$&  \textbf{\textrm{3 or 4}} \\

      (4-6) & $\alpha>0,~e^{-1}<\frac{2\gamma}{\alpha}e^{\frac{2\beta}{\alpha}}$ & $\textrm{not elementary function},~A\geq0$&  \textbf{\textrm{2}} \\

  \hline
\end{tabular}
\caption{\label{tab4} Given the scale factor $A$ equal to $A_{0}$ as $t=0$. Here $\alpha=-2c_{1},~\beta=\frac{2}{3}c_{1}-\frac{1}{2}c_{2},~\gamma=\kappa$}
\end{table}

The total of modes in the category with $w=-\frac{1}{3}$ are six and one can refer them in Table \ref{tab4}.

The two types of evolution provided by the first two modes (4-1) and (4-2) are elementary function. It's readily to
realize that there exist no $A_{I}$ satisfying $\ddot{A}_{I}=0$ in both of them, which are classified into the type 1
and the type 8 respectively.

Turn to the rest of four modes.
Observe the following equation
\begin{eqnarray}\label{a.4.3}
\gamma+\beta A^2+\alpha\ln{A}=0,
\end{eqnarray}
which will determine the allowable range of $A$ for each mode.
Its root $A_{r}$ can be expressed in terms of Lambert W function $\textbf{W}$, that is
\begin{eqnarray}\label{a.4.4}
A_{r}=\exp{[\frac{1}{2}\textbf{W}(\frac{-2\gamma}{\alpha}e^{\frac{2\beta}{\alpha}})-\frac{\beta}{\alpha}]}.
\end{eqnarray}
The Lambert W function $\textbf{W}$ includes two real branches, $\textbf{W}_{0}$ and $\textbf{W}_{-1}$\cite{LW},
which can help us fulfil the classification.

Firstly, see the mode (4-3).

Notice that when $\alpha<0$ and $\frac{-2\gamma}{\alpha}e^{\frac{2\beta}{\alpha}}\geq0$,
the principal branch $\textbf{W}_{0}>0$ is single-valued, which implies that the allowable range of $A$
locates between 0 to $A_{r}$.

On the other hand,
\begin{eqnarray}\label{a.4.5}
\nonumber A_{r}&=&\exp{[\frac{1}{2}\textbf{W}_{0}(\frac{-2\gamma}{\alpha}e^{\frac{2\beta}{\alpha}})-\frac{\beta}{\alpha}]}\\
&>&e^{-\frac{\beta}{\alpha}}=e^{\frac{1}{3}}e^{-\frac{c_{1}}{4c_{2}}}>e^{\frac{-1}{6}}e^{-\frac{c_{1}}{4c_{2}}}>0.
\end{eqnarray}
Choose $A_{I}=e^{\frac{-1}{6}}e^{-\frac{c_{1}}{4c_{2}}}$ and insert it into (\ref{2.2.3}) to obtain $\ddot{A}_{I}=0$.
This type is, different from the type 1, classified into the type 9.

Secondly, see the mode (4-4).

When $\alpha>0$ and $-\frac{1}{e}<-2\frac{\gamma}{\alpha}e^{\frac{2\beta}{\alpha}}\leq0$, the Lambert W function is double-valued, which corresponds to $\textbf{W}_{-1}<-1$ and $-1<\textbf{W}_{0}\leq0$.
The two roots of the equation (\ref{a.4.3}) are $A_{r1}=\exp{[\frac{1}{2}\textbf{W}_{-1}(\frac{-2\gamma}{\alpha}e^{\frac{2\beta}{\alpha}})-\frac{\beta}{\alpha}]}$ and $A_{r2}=\exp{[\frac{1}{2}\textbf{W}_{0}(\frac{-2\gamma}{\alpha}e^{\frac{2\beta}{\alpha}})-\frac{\beta}{\alpha}]}$.
It turns out two kinds of allowable range of $A$, $0\leq A\leq A_{r1}$ or $A\geq A_{r2}$.

When $0\leq A\leq A_{r1}$,
\begin{eqnarray}\label{a.4.6}
A_{r1}&=&\exp{[\frac{1}{2}\textbf{W}_{-1}(\frac{-2\gamma}{\alpha}e^{\frac{2\beta}{\alpha}})-\frac{\beta}{\alpha}]}\\
\nonumber&<&e^{-\frac{1}{2}-\frac{\beta}{\alpha}}=e^{\frac{-1}{6}-\frac{c_{1}}{4c_{2}}}.
\end{eqnarray}
It's readily to see $\ddot{A}_{I}=0$ for the value $A_{I}=e^{\frac{-1}{6}}e^{-\frac{c_{1}}{4c_{2}}}$, which tells us this closed universe belongs to the type 1.

When $A\geq A_{r2}$,
\begin{eqnarray}\label{a.4.6}
A_{r2}&>&e^{-\frac{1}{2}-\frac{\beta}{\alpha}}=e^{\frac{-1}{6}-\frac{c_{1}}{4c_{2}}},\\
A_{r2}&\leq&e^{-\frac{\beta}{\alpha}}=e^{\frac{1}{3}-\frac{c_{1}}{4c_{2}}}.
\end{eqnarray}
One could realize that no $A$ within the above interval will satisfy $\ddot{A}=0$. Besides, (\ref{2.2.3}) tells that
$\ddot{A}$ is positive and increasing more and more when $A$ is running far away from one. It gives us the type 3.

Thirdly, see the mode (4-5).

Given $\alpha>0$ and $-2\frac{\gamma}{\alpha}e^{\frac{2\beta}{\alpha}}=-\frac{1}{e}$, the equation (\ref{a.4.3}) has only one root $A_{r}=e^{-\frac{\beta}{\alpha}}$. In addition, all non-negative $A$ satisfy $\gamma+\beta A^2+\alpha A^2\ln{A}\geq0$. It will yield two kinds of range limit, $A\geq A_{r}$ or $0\leq A\leq A_{r}$.

For $A\geq A_{r}$, similar to the mode (4-4), one could realize that it belongs to the type 3.

For $0\leq A\leq A_{r}$,
\begin{eqnarray}\label{a.4.7}
0\leq A\leq e^{-\frac{1}{2}-\frac{\beta}{\alpha}}=A_{r},
\end{eqnarray}
which will produce $\ddot{A}_{r}=0$. Since $\ddot{A}=0$ always occurs at the maximal value of the scale factor, it implies
that the scale factor is approaching to the upper bound as the time is going toward infinity, which behaves concave downward. It will be incorporated into the type 4.

Fourthly, see the mode (4-6).

Given $\alpha>0$ and $-2\frac{\gamma}{\alpha}e^{\frac{2\beta}{\alpha}}<-\frac{1}{e}$, the equation (\ref{a.4.3}) has no real root such that the allowable range of the scale factor is $A\geq0$. Adopting the similar argument as we did for the mode (4-4), it's readily to realize that this mode affiliates to the type 2.

For $w=-\frac{1}{3}$, the effective gravitational coupling and cosmological constant are, by (\ref{1.12}),
\begin{eqnarray}
G_{eff}=\kappa+c_{1}A^2,~\Lambda_{eff}=-\frac{c_{1}}{3}-\frac{c_{2}}{2}-2c_{1}\ln{A},
\end{eqnarray}
which are the functions of time and varying with the scale factor.


\subsection{category 5: $w=0$}

Insert (\ref{3.4a}) to (\ref{2.2.1}) and get
\begin{eqnarray}\label{a.5.1}
\pm(t+t_{*})=\int\frac{\sqrt{3}\sqrt{A}}{\sqrt{\kappa+2c_{1}A^{\frac{3}{2}}-\frac{c_{2}}{2}A^3}}dA,
\end{eqnarray}
with the integration constant $t_{*}$. Both conditions $A\geq0$ and $\kappa+2c_{1}A^{\frac{3}{2}}-\frac{c_{2}}{2}A^3\geq0$ need to be satisfied. This category includes mainly twelve modes which are totally presented in terms of elementary functions. One could refer to the classification in Tables \ref{tab5}. Here we will merely examine the mode (5-5), including two
different cases in which one of them will provide a concrete example of the type 9.

Note that both of them are of the same expression of the scale factor but with different behaviors of evolution because
2nd derivative is in charge of their difference.

For the mode (5-5), given $c_{1}>0,$ $c_{2}>0$ and $2 c_{1}^2+\kappa c_{2}\geq0$, the condition $\kappa-c_{1}A^{\frac{3}{2}}+c_{2}A^{3}\geq0$ will provide the permissible range
\begin{eqnarray}
0\leq A\leq A_{r}=(\frac{2c_{1}}{c_{2}}+\frac{\sqrt{4c_{1}^2+2\kappa c_{2}}}{c_{2}})^{\frac{2}{3}}.
\end{eqnarray}

Firstly, if there exists some inflection point in the case (5-5a), by (\ref{2.2.3})
\begin{eqnarray}
A_{I}^{\frac{3}{2}}=\frac{c_{1}}{c_{2}}\pm\frac{\sqrt{c_{1}^2-4\kappa c_{2}}}{c_{2}}.
\end{eqnarray}
However, the condition $c_{1}^2\leq4\kappa c_{2}$ will force the $A_{I}$ being imaginary, which shows the lack of inflection point in the case (5-5a).

On the other hand, follow (\ref{2.2.3}) to obtain
 \begin{eqnarray}
2\frac{\ddot{A}_{r}}{A_{r}}&=&A_{r}^{\frac{-3}{2}}(c_{1}-\frac{c_{2}}{2}A^{\frac{3}{2}}_{r})\\
&=&-A_{r}^{-3}(\kappa+c_{1}A^{\frac{-3}{2}})\leq0,
\end{eqnarray}
where we've employed the relation $\kappa-c_{1}A_{r}^{\frac{3}{2}}+c_{2}A_{r}^{3}=0$, which means that the case is always concave downward. In addition, the equation (\ref{2.2.1}) tells that the mode (5-5) will experience a finite time to complete the evolution from 0 to $A_{r}$. From these consideration, one could realize the case (5-5a) is incorporated into the type 1.

Secondly, the further requirement $c_{1}^2>4\kappa c_{2}$ for the mode (5-5b) will lead to
\begin{eqnarray}
&&\frac{2c_{1}}{c_{2}}+\frac{\sqrt{4c_{1}^2+2\kappa c_{2}}}{c_{2}}
>A_{I1}=\frac{c_{1}}{c_{2}}+\frac{\sqrt{c_{1}^2-4\kappa c_{2}}}{c_{2}}\\
\nonumber&&>A_{I2}=\frac{c_{1}}{c_{2}}-\frac{\sqrt{c_{1}^2-4\kappa c_{2}}}{c_{2}}>0.
\end{eqnarray}
Since $\ddot{A}_{I1}$ and $\ddot{A}_{I2}$ are exactly vanishing, there indeed exist the flection point.
Thus, one could realize the mode (5-5b) is into the type 9, not the type 1.

Follow (\ref{1.12}) to obtain the effective gravitational coupling and cosmological constant of the category
\begin{eqnarray}
G_{eff}=\kappa+c_{1}A^{\frac{3}{2}},~\Lambda_{eff}=c_{1}A^{\frac{-3}{2}}-\frac{c_{2}}{2},
\end{eqnarray}
which are the functions of time and varying with the scale factor.

\begin{table}[tbp]\tiny
\centering
\begin{tabular}{|c|c|c|c|}
  \hline
   $w=0$ & \textrm{mode}  &  \textrm{reduced scale factor} $A$ & \textrm{type} \\  \hline

  $\begin{array}{c}\textrm{(5-1)}
   \end{array}$
   & $c_{1}=0,~ c_{2}>0$  & $A=[\sqrt{\frac{2\kappa}{c_{2}}-A^3_{0}}\sin\sqrt{\frac{3c_{2}}{8}}t+A^{\frac{3}{2}}_{0}\cos\sqrt{\frac{3c_{2}}{8}}t~]^{\frac{2}{3}},~0\leq A\leq (\frac{2\kappa}{c_{2}})^{\frac{1}{3}}$ &  \textbf{\textrm{1}} \\

  $\begin{array}{c}\textrm{(5-2)}\\
\\
\end{array}$
   & $c_{1}=0,~ c_{2}<0$ & $A=[\sqrt{\frac{2\kappa}{|c_{2}|}+A^3_{0}}\sinh\sqrt{\frac{3|c_{2}|}{8}}t+A^{\frac{3}{2}}_{0}\cosh\sqrt{\frac{3|c_{2}|}{8}}t~]^{\frac{2}{3}},~A\geq 0 $ &  \textbf{\textrm{2}} \\

  $\begin{array}{c}\textrm{(5-3)}\\
\\
   \end{array}$
   & $c_{1}>0,~ c_{2}=0$ & $A=[\frac{3}{8}c_{1}t^2+\frac{\sqrt{3}}{2}\sqrt{\kappa+2c_{1}A^{\frac{3}{2}}_{0}}t+A^{\frac{3}{2}}_{0}~]^{\frac{2}{3}},~A\geq0$&  \textbf{\textrm{2}} \\

  $\begin{array}{c}\textrm{(5-4)}\\
\\
   \end{array}$
   & $\begin{array}{c}c_{1}<0,~ c_{2}=0\end{array}$  & $A=[-\frac{3}{8}|c_{1}|t^2+\frac{\sqrt{3}}{2}\sqrt{\kappa-2|c_{1}|A^{\frac{3}{2}}_{0}}t+A^{\frac{3}{2}}_{0}~]^{\frac{2}{3}},~0\leq A\leq(\frac{\kappa}{2|c_{1}|})^{\frac{2}{3}}$&  \textbf{\textrm{1}} \\

  $\begin{array}{c}\textrm{(5-5a)}\\
\\
\\
\\
   \end{array}$
   &  $\begin{array}{c}
 2c_{1}^2+\kappa c_{2}>0,\\
c_{1}>0,~ c_{2}>0,\\
c_{1}^2\leq4\kappa c_{2}\\
\end{array}$ & $\begin{array}{c}A=[\sqrt{\frac{4c_{1}^2+2\kappa c_{2}}{c_{2}^2}-(A^{\frac{3}{2}}_{0}-\frac{2c_{1}}{c_{2}})^2}\sin\sqrt{\frac{3c_{2}}{8}}t+(A^{\frac{3}{2}}_{0}-\frac{2c_{1}}{c_{2}})\cos\sqrt{\frac{3c_{2}}{8}}t+\frac{2c_{1}}{c_{2}}~]^{\frac{2}{3}},\\
0\leq A\leq (\frac{2c_{1}+\sqrt{4c_{1}^2+2\kappa c_{2}}}{c_{2}})^{\frac{2}{3}}
\end{array}$&  \textbf{\textrm{1}} \\

  $\begin{array}{c}\textrm{(5-5b)}\\
\\
\\
\\
   \end{array}$
   &  $\begin{array}{c}2c_{1}^2+\kappa c_{2}>0,\\
c_{1}>0,~ c_{2}>0,\\
c_{1}^2>4\kappa c_{2}\\
\end{array}$ & $\begin{array}{c}A=[\sqrt{\frac{4c_{1}^2+2\kappa c_{2}}{c_{2}^2}-(A^{\frac{3}{2}}_{0}-\frac{2c_{1}}{c_{2}})^2}\sin\sqrt{\frac{3c_{2}}{8}}t+(A^{\frac{3}{2}}_{0}-\frac{2c_{1}}{c_{2}})\cos\sqrt{\frac{3c_{2}}{8}}t+\frac{2c_{1}}{c_{2}}~]^{\frac{2}{3}},\\
0\leq A\leq (\frac{2c_{1}+\sqrt{4c_{1}^2+2\kappa c_{2}}}{c_{2}})^{\frac{2}{3}}
\end{array}$&  \textbf{\textrm{9}} \\

  $\begin{array}{c}\textrm{(5-6)}\\
\\
\\
   \end{array}$
   & $\begin{array}{c}2c_{1}^2+\kappa c_{2}>0,\\
c_{1}>0,~ c_{2}<0\\
\end{array}$  & $A=[\sqrt{\frac{2}{|c_{2}|}}\sqrt{\kappa+2c_{1}A^{\frac{3}{2}}_{0}+\frac{|c_{2}|}{2}A^3_{0}}\sinh\sqrt{\frac{3|c_{2}|}{8}}t
+\frac{2c_{1}+|c_{2}|A^{\frac{3}{2}}_{0}}{|c_{2}|}\cosh\sqrt{\frac{3|c_{2}|}{8}}t-\frac{2c_{1}}{|c_{2}|}~]^{\frac{2}{3}},~A\geq 0 $& \textbf{\textrm{2}}  \\

 $\begin{array}{c}\textrm{(5-7)}\\
\\
\\
\end{array}$
  & $\begin{array}{c}2c_{1}^2+\kappa c_{2}>0,\\
c_{1}<0,~ c_{2}>0\\
\end{array}$  & $\begin{array}{c}A=[\sqrt{\frac{4|c_{1}|^2+2\kappa c_{2}}{c_{2}^2}-(A^{\frac{3}{2}}_{0}+\frac{2|c_{1}|}{c_{2}})^2}\sin\sqrt{\frac{3c_{2}}{8}}t
+(A^{\frac{3}{2}}_{0}+\frac{2|c_{1}|}{c_{2}})\cos\sqrt{\frac{3c_{2}}{8}}t-\frac{2|c_{1}|}{c_{2}}~]^{\frac{2}{3}},\\
0\leq A\leq (\frac{-2|c_{1}|+\sqrt{4|c_{1}|^2+2\kappa c_{2}}}{c_{2}})^{\frac{2}{3}}\end{array}$ & \textbf{\textrm{1}}  \\

  $\begin{array}{c}\textrm{(5-8)} \\
  \\
\\
  \end{array}$

  & $\begin{array}{c}
  2c_{1}^2+\kappa c_{2}>0,\\
c_{1}<0,~ c_{2}<0\\
  \end{array}$  &

  $\begin{array}{c}
  A=[\sqrt{\frac{2}{|c_{2}|}}\sqrt{\kappa-2|c_{1}|A^{\frac{3}{2}}_{0}+\frac{|c_{2}|}{2}A^3_{0}}\sinh\sqrt{\frac{3|c_{2}|}{8}}t
+\frac{-2|c_{1}|+|c_{2}|A^{\frac{3}{2}}_{0}}{|c_{2}|}\cosh\sqrt{\frac{3|c_{2}|}{8}}t+\frac{2|c_{1}|}{|c_{2}|}~]^{\frac{2}{3}},\\
0\leq A\leq (\frac{2|c_{1}|-\sqrt{4|c_{1}|^2-2\kappa |c_{2}|}}{|c_{2}|})^{\frac{2}{3}}~\textrm{or}~A\geq (\frac{2|c_{1}|+\sqrt{4|c_{1}|^2-2\kappa |c_{2}|}}{|c_{2}|})^{\frac{2}{3}}
\end{array}$ & $\begin{array}{l}\textbf{\textrm{1 or 3}}
\end{array}$
  \\

$\begin{array}{l}\textrm{(5-9)}\\
\\
 \end{array}$
 &
$\begin{array}{c}2c_{1}^2+\kappa c_{2}=0,\\
c_{1}>0,~c_{2}<0\\
\end{array}$
   &
    $A=[(A^{\frac{3}{2}}_{0}+\frac{\kappa}{c_{1}})\exp{(\sqrt{\frac{3}{\kappa}}\frac{c_{1}}{2}~t)}-\frac{\kappa}{c_{1}}]^{\frac{2}{3}},~A\geq0$
     &  \textbf{\textrm{2}} \\

$\begin{array}{c}
\textrm{(5-10)}\\
\\
\end{array}$
 &

$\begin{array}{c}
2c_{1}^2+\kappa c_{2}=0,\\
c_{1}<0,~ c_{2}<0\\
\end{array}$
&
$\begin{array}{l}
A=[(A^{\frac{3}{2}}_{0}-\frac{\kappa}{|c_{1}|})\exp{(\sqrt{\frac{3}{\kappa}}\frac{|c_{1}|}{2}~t)}+\frac{\kappa}{|c_{1}|}]^{\frac{2}{3}},~
A\geq(\frac{\kappa}{|c_{1}|})^{\frac{2}{3}}~\textrm{or}~0\leq A\leq(\frac{\kappa}{|c_{1}|})^{\frac{2}{3}}
\end{array}$
&  $\begin{array}{l}\textbf{\textrm{3 or 4}}
\end{array}$ \\

$\begin{array}{l}
\textrm{(5-11)}\\
\\
\end{array}$
&
$\begin{array}{l}
2c_{1}^2+\kappa c_{2}<0,\\
c_{1}>0,~ c_{2}<0\\
\end{array}$  & $A=[\sqrt{\frac{2}{|c_{2}|}}\sqrt{\kappa+2c_{1}A^{\frac{3}{2}}_{0}+\frac{|c_{2}|}{2}A^3_{0}}\sinh\sqrt{\frac{3|c_{2}|}{8}}t
+\frac{2c_{1}+|c_{2}|A^{\frac{3}{2}}_{0}}{|c_{2}|}\cosh\sqrt{\frac{3|c_{2}|}{8}}t-\frac{2c_{1}}{|c_{2}|}~]^{\frac{2}{3}},~A\geq 0$ &  \textbf{\textrm{2}} \\

$\begin{array}{l}
\textrm{(5-12)}\\
\\
\end{array}$
&
$\begin{array}{l}
2c_{1}^2+\kappa c_{2}<0,\\
c_{1}<0,~ c_{2}<0\\
\end{array}$
&
 $A=[\sqrt{\frac{2}{|c_{2}|}}\sqrt{\kappa-2|c_{1}|A^{\frac{3}{2}}_{0}+\frac{|c_{2}|}{2}A^3_{0}}\sinh\sqrt{\frac{3|c_{2}|}{8}}t
+\frac{-2|c_{1}|+|c_{2}|A^{\frac{3}{2}}_{0}}{|c_{2}|}\cosh\sqrt{\frac{3|c_{2}|}{8}}t+\frac{2|c_{1}|}{|c_{2}|}~]^{\frac{2}{3}},~A\geq0$ &  \textbf{\textrm{2}} \\
\hline
\end{tabular}
\caption{\label{tab5} Given the scale factor $A$ equal to $A_{0}$ as $t=0$. }
\end{table}


\section{Numerical simulation}

There are two ways to cause the accelerated expansion of the universe. One is to introduce an exotic matter field with enough negative pressure and the other is to modify the geometry. We want to examine that the if $f(R,T)$ gravity with normal matter field is afford to explain the accelerated expansion of the universe instead of imposing an exotic matter field into Einstein's gravity. Since nowadays the pressure of normal matter can be ignored in our universe, we will focus the modes if the category 5 which describes the behavior of dust can achieve our target.

\begin{figure}
  \begin{minipage}[tbp]{0.5\linewidth}
    \centering
    \includegraphics[scale=0.3]{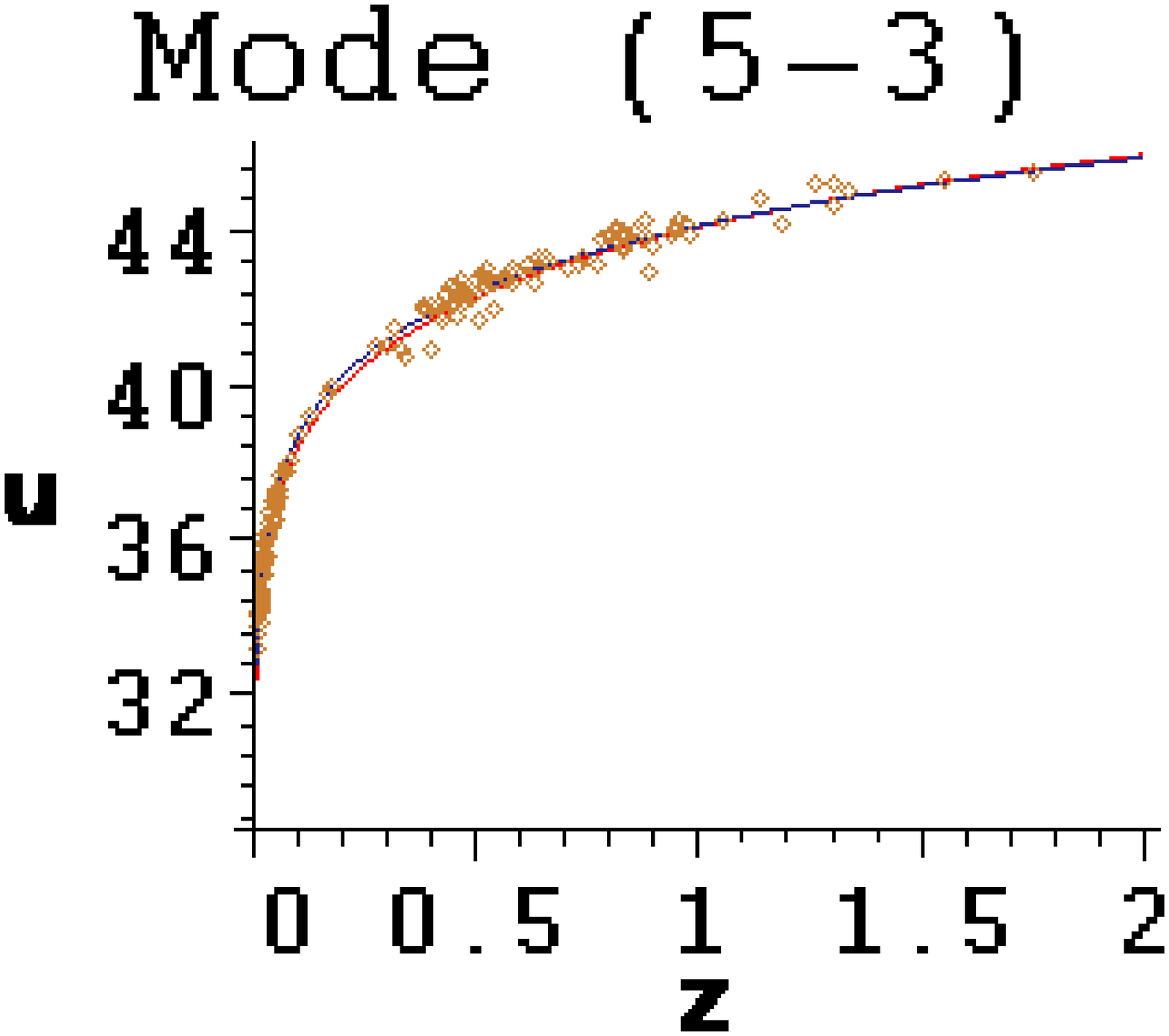}
  \caption{\label{fig7}Here shows the relation between
  \protect\\ the distance modulus $\mu$ and the
  redshift $z$.
 \protect\\ The result of the mode (5-3) with $c_{1}=0.55$
  \protect\\ is plotted by the blue line. The supernovae
  \protect\\ data points, plotted with yellow points,
  \protect\\ come from \cite{Riess04}. The result of the standard
  \protect\\ $\Lambda$CDM model ($m = 0.3$,$M = 0.7$) is plotted
  \protect\\ by the red line.}
  \end{minipage}%
  \begin{minipage}[tbp]{0.5\linewidth}
    \centering
    \includegraphics[scale=0.3]{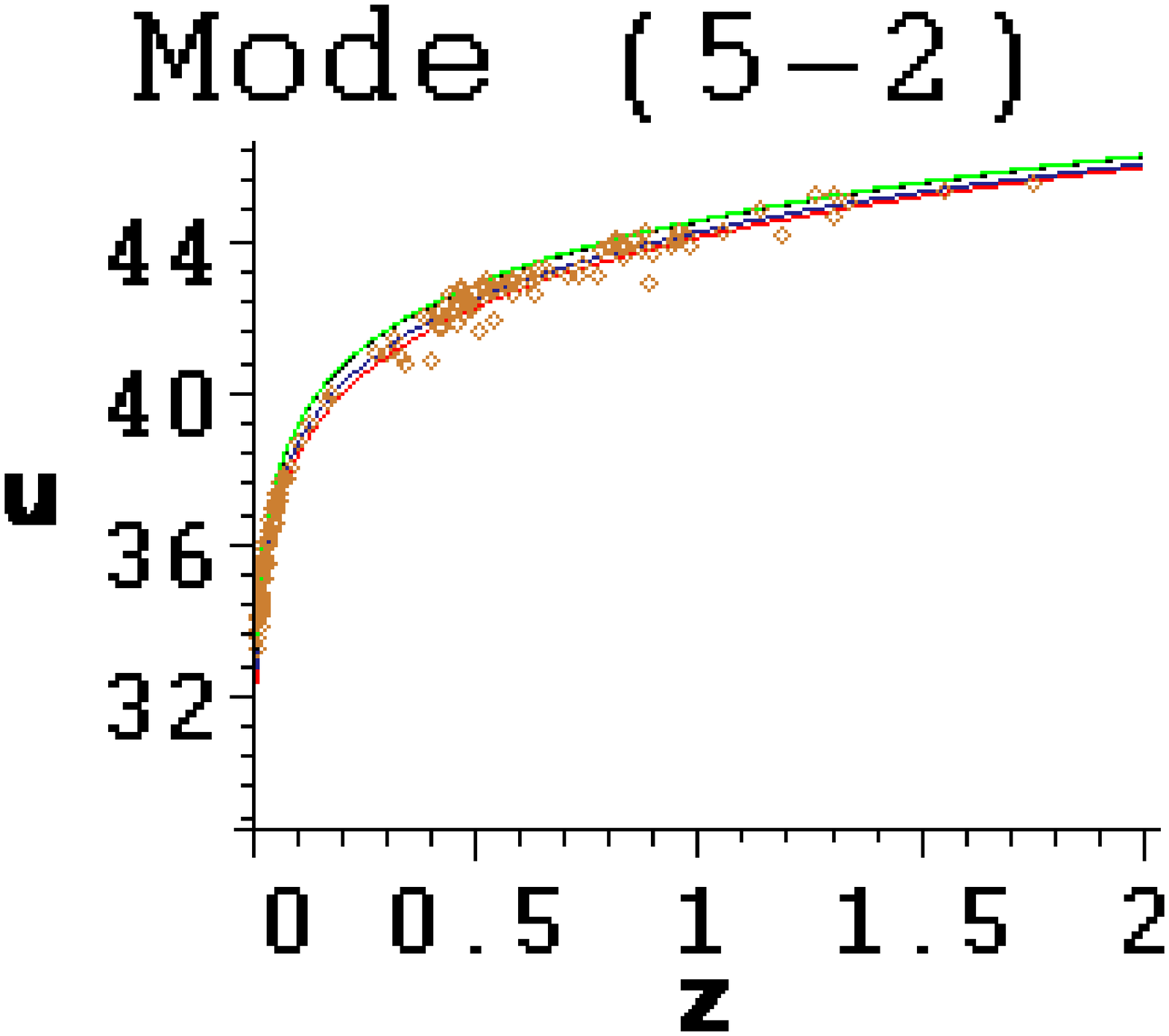}
  \caption{\label{fig8}Here shows the relation between the distance modulus $\mu$ and the
redshift $z$. The result of the mode (5-2) with $|c_{2}|=2.5$ is plotted by the blue line. The supernovae data points, plotted with yellow points, come from \cite{Riess04}. The
result of the standard $\Lambda$CDM model ($m = 0.3$,
$M = 0.7$) is plotted by the red
line. The lines (the green line with $c_{1}=0.61$ and the black line with $c_{1}=0.76$) with the $c_{2}$ fitting the range \ref{red} lies almost away from the observational data.}
  \end{minipage}
    \begin{minipage}[tbp]{0.5\linewidth}
    \centering
  \includegraphics[scale=0.3]{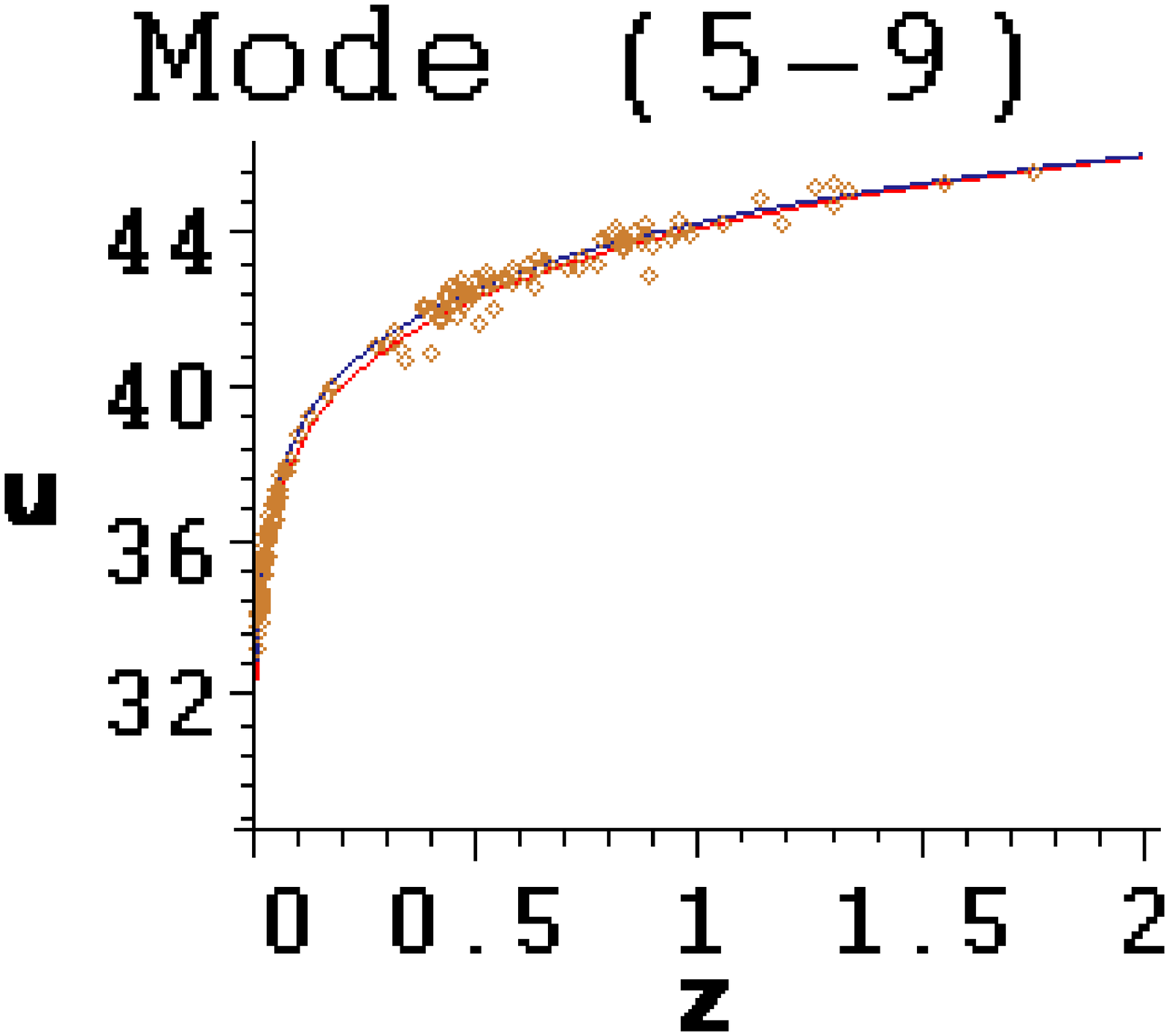}\\
  \vspace*{8pt}
  \caption{\label{fig9}Here shows the relation between the distance modulus $\mu$ and the
redshift $z$. The result of the mode (5-9) with $c_{1}=0.37$ is plotted by the blue line. The supernovae data points, plotted with yellow points, come from \cite{Riess04}. The
result of the standard $\Lambda$CDM model ( $m = 0.3$,
$M = 0.7$) is plotted by the red
line.}
  \end{minipage}
\end{figure}

Riess et al.\cite{Riess04} discovered 16 Type Ia supernovae and showed the first conclusive evidence for cosmic deceleration that preceded the current epoch of cosmic acceleration. Thus, we will pay attention to the type 2 which is only type with a transition from deceleration to acceleration in a period of expansion. In the following, we will look into the three modes \textrm{(5-3)}, \textrm{(5-2)} and \textrm{(5-9)} which are only involved in one parameter $c_{1}$ or $c_{2}$ so that are easier to analysis than other modes.
The relation between distance modulus versus redshift of the three modes are plotted in Figure \ref{fig7},\ref{fig8},\ref{fig9}.

Define the redshift
\begin{eqnarray}
z=\frac{A(1)}{A(t)}-1
\end{eqnarray}
and the luminosity distance for FLRW spacetime with $K=0$\cite{Cope,Riess04}
\begin{eqnarray}
d_{L}=\frac{c}{H_{0}}(1+z)\int^{1}_{t}\frac{dt}{a(t)}=\frac{c}{H_{0}}(1+z)\int^{z}_{0}\frac{dz}{H(z)},
\end{eqnarray}
whose unit is megaparsec. Here the present Hubble constant $H_{0}=70\frac{Km}{sec-Mpc}$.

The distance modulus $\mu$ is estimated by the difference between the flux (apparent magnitude) $m$ and the luminosity (absolute magnitude) $M$,
that is\cite{Cope,Riess04},
\begin{eqnarray}
\mu=m-M=5\log_{10}{\frac{d_{L}}{Mpc}}+25.
\end{eqnarray}

For comparison, we also plot the prediction of the $\Lambda$CDM model\cite{Cope,Riess04} and the astronomical observation data\cite{Riess04}.
The luminosity distance in $\Lambda$CDM model is defined as
\begin{eqnarray}
d_{L}=\frac{c}{H_{0}}(1+z)\int^{z}_{0}\frac{dz}{\sqrt{(1+z)^2(1+\Omega_{m}z)-z(2+z)\Omega_{\Lambda}}},
\end{eqnarray}
with the standard values of the parameters $\Omega_{m}= 0.3$ and $\Omega_{\Lambda} = 0.7$.
The plots show that the mode (5-2), (5-3) and (5-9) match the astronomical observational data well. They give an accelerating universe just like the $\Lambda$CDM model does for small redshift $z$ (e.g., $z<2$).

According to the model-independent estimate\cite{Riess04}, the decelerated-accelerated transition occurs at $z_{t}=0.46\pm0.13$. One could estimate the range of the  parameters $c_{1}$ and $c_{2}$ in order to see if the three modes are compatible with the $z_{t}$. It turns out
\begin{eqnarray}\label{red}
\nonumber&&\textrm{mode (5-2):}~~~0.61\leq |c_{2}|\leq 0.76\\
\nonumber&&\textrm{mode (5-3):}~~~0.53\leq c_{1}\leq 0.81\\
&&\textrm{mode (5-9):}~~~0.33\leq c_{1}\leq 0.41.
\end{eqnarray}
Based on the consideration on $z_{t}$, the modes \textrm{(5-2)} and \textrm{(5-9)} could be better than the mode \textrm{(5-3)} because the latter has no a good fitting relation between the distance modulus versus the redshift as the parameter $c_{1}$ is within the above range.

\section{Conclusion}

In Einstein's gravity, the accelerated expansion of the universe is attributed to the cosmological constant or the contribution of an exotic matter field that is equivalent to a kind of scalar field with the specific potential. Instead of introducing an exotic matter field we examine the $f(R,T)$ gravity, a kind of modified gravity with non-minimal coupling to matter field in this article.

For the sake of simplification and definiteness, we consider the specific form of $f(R,T)=R+F(T)$ and employ the assumption of adiabatic condition to look for the characterized pattern of $F(T)$. We've seen that $f(R,T)$ cosmology with $K=0$ contains nine different types of cosmological evolution whose major distinctions are listed in  Table 5. One could also refer to the exact expression for each mode in Table 1, 2, 3, 4, except for the modes (4-3)--(4-6) which are not able to expressed by elementary functions. The type 1 and 9 are displaying two different closed universe that are absent in standard FLRW cosmology with $K=0$. The type 1 is decelerated expanding first, reaching the maximal size, then accelerated contracting, which will result in the big crunch inevitably.
The type 9 is also the type of the big brunch, but it will experience a period of accelerated expansion before reaching the maximal scale and decelerated contraction before the big crunch. The type 5, static universe, is too trivial to ignore it. The type 4 is upper bound, starting from the singularity and always expanding in a decelerated rate up to a maximal scale. Other five are open types which are all involved in accelerated expansion.

To observe luminosity distances of high redshift supernovae will provide the direct evidence of current acceleration of the universe. We compare the three dust modes \textrm{(5-2)}, \textrm{(5-3)} and \textrm{(5-9)} with the supernovae data and obtain the good data fitting. Further, after considering the allowable range of the redshift $z_{t}$ corresponding to the transition, there are still two modes \textrm{(5-2)} and \textrm{(5-9)} good enough to fit data.
In addition, note that there is merely one parameter $c_{1}$ or $c_{2}$ needed in our simulation, which is the advantage of our model. Even if we adopt other three modes \textrm{(5-6)}, \textrm{(5-11)}, \textrm{(5-12)}, we need to control two parameters at most.

Incidentally, we use the equation (\ref{1.13}) rather than the original equation (15) used in \cite{Harko}, even if we both operate the covariant derivative on the field equations coming from the variation of the Lagrangian in order to get what we need.
If we limit ourselves to original settings, some easy calculations show the operation of the equations (\ref{2.2.1}) and (\ref{2.2.2}) will lead to the equation (\ref{2.3}). This is not surprising. Like solving the standard FLRW equations, there are only two independent equations. Any other equation coming from the Bianchi identity is nothing but a combination of the two independent equations. Nevertheless, if we adopt the equation (15) used in \cite{Harko}, it turns out
\begin{eqnarray}\label{6.1}
\frac{d}{dt}[a^3(\rho+p)(\kappa+F_{T})]-\kappa a^3\dot{p}=0.
\end{eqnarray}
To integrate the above equation will get
\begin{eqnarray}\label{2.4}
\dot{H}+\frac{3}{2}H^2+\frac{K}{2a^2}+\frac{\kappa w}{2}\rho=\mu,
\end{eqnarray}
with $\mu$ a constant. In fact, it is just the equation (\ref{2.2.2}) with $F=2\mu$. Hence $F(T)$ is always a constant  such that a lot of possible modes are ignored. That is why we adopt our setting in order for generality.\\


\end{document}